\documentclass[12pt, draftcls, onecolumn]{IEEEtran}  

\IEEEoverridecommandlockouts                              
\usepackage{graphicx}
\usepackage{graphics}                                     
\usepackage{amsmath,amssymb}
\usepackage{mathrsfs}
\usepackage{multicol}
\usepackage{subfigure}
\title{\Huge
Generalized Nonlinear Robust Energy-to-Peak Filtering for
Differential Algebraic Systems}

\author{Masoud Abbaszadeh
\thanks{{\tt\small e-mail: masoud@ualberta.net}}%
}

\begin{document}

\maketitle \thispagestyle{empty} \pagestyle{empty}

\begin{abstract}
The problem of robust nonlinear energy-to-peak filtering for nonlinear descriptor systems
with model uncertainties is addressed.
The system is assumed to have nonlinearities both in the state and output equations
as well as norm-bounded time-varying uncertainties in the realization matrices.
A generalized nonlinear dynamic filtering structure is proposed
for such a class of systems with more degrees of freedom
than the conventional static-gain and dynamic filtering structures.
The $\mathcal{L}_{2}-\mathcal{L}_{\infty}$ filter is
synthesized through semidefinite programming and strict LMIs, in which the
energy-to-peak filtering performance in optimized.
\end{abstract}

\emph{keywords}: Robust Filtering, Energy-to-Peak Filtering,
Nonlinear $\mathcal{L}_{2}-\mathcal{L}_{\infty}$, Descriptor Systems,
Semidefinite Programming, Lipschitz Systems, DAE

\section{Introduction}

\PARstart{D}{escriptor} systems, also referred to as singular systems or differential-algebraic
equation (DAE) systems, arise from an inherent and natural modeling
approach, and have vast applications in engineering disciplines such as
power systems, network and circuit analysis, and multibody mechanical systems,
as well as in social and economic sciences. Generalizing the regular state space modeling
(i.e. pure ODE systems), descriptor systems can characterize a larger class of
systems than conventional state space models
and can \emph{describe} the physics of the system more precisely.
Many approaches have been developed to design state
observers for descriptor systems.
Observer design and filtering of nonlinear dynamic
systems has been a subject of extensive research in the last decade
due to its theoretical and practical importance.
In \cite{Dai, Boutayeb1, Hou, Darouch1, Darouch2, Uezato, Lu_G3, He_Wang, Masubuchi,
Shields, Zimmer} various methods of observer design for linear and
nonlinear descriptor systems have been proposed. In \cite{Boutayeb1}
an observer design procedure is proposed for a class nonlinear
descriptor systems using an appropriate coordinate transformation.
In \cite{Shields}, the authors address the unknown input observer
design problem dividing the system into two dynamic and static
subsystems. Reference \cite{Lu_G3} studies the full order and
reduced order observer design for Lipschitz nonlinear Systems.

A fundamental limitation encountered in conventional observer theory
is that it can not guarantee observer performance in the presence of model uncertainties
and/or disturbances and measurement noise.
One of the most popular ways to deal with the nonlinear state estimation problem is the extended Kalman
filtering. However, the requirements of specific noise statistics and weakly nonlinear dynamics, has restricted its
applicability to nonlinear systems. To deal with the nonlinear filtering problem in
the presence of model uncertainties and unknown exogenous disturbances, we can resort the
\emph{robust} $H_{\infty}$ filtering, and $\mathcal{L}_{2}-\mathcal{L}_{\infty}$ filtering approaches.
See for example \cite{deSouza1,deSouza2,Meng.etal2007,Palhares.etal2000,Qiu.etal2008,Abbaszadeh3,Abbaszadeh4, Abbaszadeh5,abbaszadeh2008lmi,abbaszadeh2012generalized}
and the references therein.
The mathematical system model is assumed to be affected by
time-varying parametric uncertainties, while norm bounded disturbances affect the measurements.
Each of the two criteria has its own physical implications and applications.
In $H_{\infty}$ filtering, the $\mathcal{L}_{2}$ gain from the exogenous disturbance to the filter error is guaranteed
to be less than a prespecified level. Therefore, this $\mathcal{L}_{2}$ gain minimization is in fact an \emph{energy-to-energy}
filtering problem. In $\mathcal{L}_{2}-\mathcal{L}_{\infty}$ filtering,
the ratio of the peak value of the error ($\mathcal{L}_{\infty}$ norm)
to the energy of disturbance ($\mathcal{L}_{2}$ norm) is considered, therefore, conforming an \emph{energy-to-peak} performance.
This strategy has been used for both full-order and reduced-order filter design
through LMIs in \cite{Grigoriadis.etal1997_1, Gao.etal2003_1} and also as a way for model reduction
in \cite{Grigoriadis.etal1997_2}.
Recently, $\mathcal{L}_{2}-\mathcal{L}_{\infty}$ filtering has been
addressed for linear descriptor systems \cite{Zhou.etal2008_1,Zhang.etal2010_1}.
However, the problem of $\mathcal{L}_{2}-\mathcal{L}_{\infty}$ filtering for
nonlinear descriptor systems has not been fully investigated yet, despite the practical
motivation and the great importance.

In this paper, we study the \emph{robust} nonlinear
$\mathcal{L}_{2}-\mathcal{L}_{\infty}$ filtering
for continuous-time Lipschitz descriptor systems in the presence of
disturbance and model uncertainties, in the LMI optimization framework.
We consider nonlinearities in both the state and output equations,
Furthermore, we generalize the filter structure by
proposing a general dynamical filtering framework that can easily
capture both dynamic and static-gain filter structures as special cases.
The proposed dynamical structure has additional degrees of freedom compared to
conventional static-gain filters and consequently is capable of
robustly stabilizing the filter error dynamics for
systems for which an static-gain filter can not be found.

Stability of nonlinear ODE systems is established through Lyapunov theory,
while the stability of DAE systems is established through LaSalle's invariant set theory.
The results on ODEs, such as in
\cite{Meng.etal2007,Palhares.etal2000,Qiu.etal2008,Abbaszadeh3,Abbaszadeh4, Abbaszadeh5},
are directly cast into strict LMIs while the results here are a set of linear matrix equations
and inequalities leading into a semidefinite programming.
The developed SDP problem is then smartly converted into a
strict LMI formulation, without any approximations, and is
efficiently solvable by readily available LMI solvers.

The rest of the paper is organized as follows. In section II, the problem statement and some
preliminaries are mentioned. In section III, we propose a new method
for robust $\mathcal{L}_{2}-\mathcal{L}_{\infty}$ filter design for nonlinear descriptor
uncertain systems based on semidefinite programming (SDP).
In Section IV, the SDP problem of Section III is
converted into strict LMIs. In section V, we show the proposed filter
design procedure through an illustrative example.


\section{Preliminaries and Problem Statement}

Consider the following class of continuous-time uncertain nonlinear
descriptor systems:
\begin{align}
\left(\Sigma_{s} \right): \mathbf{E}\dot{x}(t)&=(A+\Delta
A(t))x(t)+\Phi(x,u)+B w(t)\label{sys1}
\\ y(t)&=(C+\Delta C(t))x(t)+\Psi(x,u)+Dw(t)\label{sys2}
\end{align}
where $x\in {\mathbb R} ^{n} ,u\in {\mathbb R} ^{m} ,y\in {\mathbb
R} ^{p} $ and $\Phi(x,u)$ and $\Psi(x,u)$ contain nonlinearities of
second order or higher. $\mathbf{E}$, $A$, $B$, $C$ and $D$ are
constant matrices with compatible dimensions; $\mathbf{E}$ may be
singular. When the matrix $\mathbf{E}$ is singular, the above form is equivalent to a set of
differential-algebraic equations (DAEs) \cite{Dai}.
In other words, the dynamics of descriptor systems, comprise a set of differential
equations together with a set of algebraic constraints. Unlike conventional
state space systems in which the initial conditions can be
freely chosen in the operating region, in the descriptor systems,
initial conditions must be \emph{consistent}, i.e.
they should satisfy the algebraic constraints.
Consistent initialization of descriptor systems naturally happens in physical
systems but should be taken into account when simulating such systems \cite{Pantelides}.
Without loss of generality, we assume that
$0<rank(\mathbf{E})=s<n$; $x(0)=x_{0}$ is a consistent (unknown) set of initial conditions.
If the matrix $\mathbf{E}$ is non-singular (i.e. full rank), then the descriptor form reduces to the conventional state space.
The number of algebraic constraints that must be satisfied by $x_{0}$ equals $n-s$.
We assume the pair $(\mathbf{E},A)$ to be \emph{regular},
i.e. $\det(s\mathbf{E}-A) \neq 0$ for some $s \in \mathbb{C}$ and $(\mathbf{E},A,C)$
to be observable, i.e. \cite{Ishihara}
\begin{align}
rank \left[
       \begin{array}{c}
         s\mathbf{E}-A \\
         C \\
       \end{array}
     \right]=n, \ \forall \ s \in \mathbb{C}.\notag
\end{align}
We also assume that the system (\ref{sys1})-\eqref{sys2} is locally Lipschitz with
respect to $x$ in a region $\mathcal{D}$ containing the origin,
uniformly in $u$, i.e.:
\begin{align}
&\Phi(0,u^{*})=\Psi(0,u^{*})=0,\notag\\
&\|\Phi(x_{1},u^{*})-\Phi(x_{2},u^{*})\|\leqslant\gamma_{1}\|x_{1}-x_{2}\|
,\hspace{2mm} \forall \, x_{1},x_{2}\in \mathcal{D}\notag\\
&\|\Psi(x_{1},u^{*})-\Psi(x_{2},u^{*})\|\leqslant\gamma_{2}\|x_{1}-x_{2}\|
,\hspace{2mm} \forall \, x_{1},x_{2}\in \mathcal{D}\notag
\end{align}
where $\|.\|$ is the induced 2-norm, $u^{*}$ is any admissible
control signal and $\gamma_{1}, \ \gamma_{2}>0$ are the Lipschitz constants
of $\Phi(x,u)$ and $\Psi(x,u)$, respectively. If the nonlinear functions
$\Phi(x,u)$ and $\Psi(x,u)$ satisfy the Lipschitz continuity
condition globally in $\mathbb{R}^{n}$, then the results will be
valid globally. $w(t)\in\mathcal{L}_{2}[0,\infty)$ is an unknown
exogenous disturbance, and $\Delta A(t)$ and $\Delta C(t)$ are
unknown matrices representing time-varying parameter uncertainties,
and are assumed to be of the form
\begin{eqnarray}
\left[
  \begin{array}{c}
    \Delta A(t) \\
    \Delta C(t) \\
  \end{array}
\right]= \left[
  \begin{array}{c}
    M_{1} \\
    M_{2} \\
  \end{array}
\right]F(t)N \label{uncer1}
\end{eqnarray}
where $M_{1}$, $M_{2}$ and $N$ are known real constant matrices and
$F(t)$ is an unknown real-valued time-varying matrix satisfying
\begin{equation}
F^{T}(t)F(t)\leq I \hspace{1cm} \forall \ t\in [0,\infty).
\end{equation}
The parameter uncertainty in the linear terms can be regarded as the
variation of the operating point of the nonlinear system. It is also
worth noting that the structure of parameter uncertainties in
(\ref{uncer1}) has been widely used in the problems of robust
control and robust filtering for both continuous-time and
discrete-time systems and can capture the uncertainty in a number of
practical situations \cite{deSouza1}, \cite{Khargonekar}.

\subsection{Filter Structure}
We propose the general filtering framework of the following form
\begin{equation}
\begin{split}
\left(\Sigma_{o} \right):\mathbf{E}\dot{x}_{F}(t)&=
A_{F}x_{F}(t)+B_{F}y(t)+\mathcal{E}_{1}\Phi(x_{F},u)+\mathcal{E}_{2}\Psi(x_{F},u)\\
z_{F}(t)&=C_{F}x_{F}(t)+\mathcal{E}_{3}\Psi(x_{F},u).\label{observer1}
\end{split}
\end{equation}
The proposed framework can capture both dynamic and static-gain
filter structures by proper selection of $\mathcal{E}_{1}$, $\mathcal{E}_{2}$ and
$\mathcal{E}_{3}$. Choosing $\mathcal{E}_{1}=I$, $\mathcal{E}_{2}=0$ and
$\mathcal{E}_{3}=0$ leads to the following dynamic filter
structure:
\begin{equation}
\begin{split}
\mathbf{E}\dot{x}_{F}(t)&=
A_{F}x_{F}(t)+B_{F}y(t)+\Phi(x_{F},u)\\
z_{F}(t)&=C_{F}x_{F}(t).\label{observer2}
\end{split}
\end{equation}
Furthermore, for the static-gain filter structure we have:
\begin{equation}
\begin{split}
\mathbf{E}\dot{x}_{F}(t)&=Ax_{F}(t)+\Phi(x_{F},u)+L[y(t)-Cx_{F}(t)-\Psi(x_{F},u)]\\
z_{F}(t)&=x_{F}(t).\label{observer3}
\end{split}
\end{equation}
Hence, with
$A_{F}=A-LC, \ B_{F}=L, \ C_{F}=I, \mathcal{E}_{1}=I, \ \mathcal{E}_{2}=-L, \ \mathcal{E}_{3}=0$,
the general filter captures the well-known static-gain observer filter
structure as a special case. We prove our result for the general
filter of class $(\Sigma_{o})$.

Now, suppose $z(t)=Hx(t)$ stands for the controlled output for states to be estimated where
$H$ is a known matrix. The estimation error is defined as
\begin{align}
e(t)\triangleq z(t)-z_{F}(t)=-C_{F}x_{F}+(H-C)x-\mathcal{E}_{3}\Psi(x_{F},u).\label{error1}
\end{align}
The filter error dynamics is given by
\begin{align}
\left(\Sigma_{e}
\right):\mathbf{\widetilde{E}}\dot{\xi}(t)&=(\widetilde{A}+\Delta
\widetilde{A})\xi(t)+S_{1}\Omega(\xi,u)+\widetilde{B}w(t)\\
e(t)&=\widetilde{C}\xi(t)+S_{2}\Omega(\xi,u),
\end{align}
where,
\begin{align}
&\xi\triangleq\left[
       \begin{array}{c}
         x_{F} \\
         x \\
       \end{array}
     \right], \widetilde{A}=\left[
     \begin{array}{cc}
       A_{F} & B_{F}C \\
       0 & A \\
     \end{array}
   \right], \Delta \widetilde{A}=\left[
                                   \begin{array}{cc}
                                     0 & B_{F}\Delta C \\
                                     0 & \Delta A \\
                                   \end{array}
                                 \right]\notag\\
&\mathbf{\widetilde{E}}=\left[
                          \begin{array}{cc}
                            \mathbf{E} & 0 \\
                            0 & \mathbf{E} \\
                          \end{array}
                        \right]
, \widetilde{B}=\left[
                                \begin{array}{c}
                                  B_{F}D \\
                                  B \\
                                \end{array}
                              \right], \widetilde{C}=\left[
     \begin{array}{cc}
       -C_{F} & H \\
     \end{array}
   \right], \notag\\
&\Omega(\xi,u)=\left[
                            \begin{array}{cccc}
                                \Phi(x,u) & \Psi(x,u) & \Phi(x_{F},u) & \Psi(x_{F},u)
                            \end{array}
                      \right]^{T}\notag\\
&S_{1}=\left[
         \begin{array}{cccc}
           0 & B_{F} & \mathcal{E}_{1} & \mathcal{E}_{2} \\
           I & 0     & 0 & 0 \\
         \end{array}
       \right], S_{2}=\left[
                        \begin{array}{cccc}
                          0 & 0 & 0 & -\mathcal{E}_{3} \\
                        \end{array}
                      \right] \notag.
\end{align}
For the nonlinear function $\Omega$, it is easy to show that
\begin{align}
&\Gamma\triangleq\left[
                   \begin{array}{cccc}
                     0          & 0          & \gamma_{1} & \gamma_{2}\\
                     \gamma_{1} & \gamma_{2} & 0          & 0\\
                   \end{array}
                 \right]^{T}, \ \|\Gamma\|\ = \sqrt{\gamma_{1}^{2}+\gamma_{2}^{2}} \\
&\|\Omega(\xi_{1},u)-\Omega(\xi_{2},u)\|\leq\|\Gamma(\xi_{1}-\xi_{2})\| =\sqrt{\gamma_{1}^{2}+\gamma_{2}^{2}}\|\xi_{1}-\xi_{2}\|\triangleq
\gamma\|\xi_{1}-\xi_{2}\|\label{Gamma1}.
\end{align}
Thus, the filter error system is Lipschitz with Lipschitz constant
$\gamma$.

\subsection{Disturbance Attenuation Level}

Our purpose is to design the filter matrices $A_{F}$, $B_{F}$ and $C_{F}$,
such that in the absence of disturbance, the filter error dynamics is
asymptotically stable and moreover, for all $w(t) \in \mathcal{L}_{2}[0,\infty)$, subject to
zero error initial conditions, the following $\mathcal{L}_{2}-\mathcal{L}_{\infty}$ norm upper bound is
simultaneously guaranteed.
\begin{equation}
\|e\|_{\infty} \leq \mu \|w\|_{2},
\end{equation}
where $\|.\|_{\infty}$ and $\|.\|_{2}$ denote the signal $2-norm$ and $infinity-norm$,
respectively, defined as:
\begin{align}
\|w(t)\|_{2} &= \sqrt{\int^{\infty}_{0}\left(w^{T}(t)w(t)\right)dt}\notag\\
\|e(t)\|_{\infty} &= \sup_{t} \sqrt{|e(t)|^{2}} \ \ \ \forall \ t \in [0,\infty).\notag
\end{align}

In the following, we mention some useful lemmas that will be used
later in the proof of our results. \\

\emph{\textbf{Lemma 1. \cite{deSouza2}} For any
$x,y\in\mathbb{R}^{n}$ and any positive definite matrix
$P\in\mathbb{R}^{n\times{n}}$, we have}
\begin{equation}
2x^{T}y\leq x^{T}Px+y^{T}P^{-1}y.\notag
\end{equation}

\emph{\textbf{Lemma 2. \cite{deSouza2}} Let $A,D, E, F$ and P be
real matrices of appropriate dimensions with $P>0$ and $F$
satisfying $F^{T}F\leq I$. Then for any scalar $\epsilon>0$
satisfying $P^{-1}-\epsilon^{-1}DD^{T}>0$, we have}
\begin{align}
(A+DFE)^{T}P(A+DFE)\leq A^{T}(P^{-1}-\epsilon^{-1}DD^{T})^{-1}A+\epsilon E^{T}E.\notag
\end{align}

\emph{\textbf{Lemma 3. \cite[p. 301]{Horn1}} A matrix $A \in
\mathbb{R}^{n \times n}$ is invertible if there is a matrix norm
$\||.\||$ such that $\||I-A\||<1$.}


\section{$\mathcal{L}_{2}-\mathcal{L}_{\infty}$ Filter Synthesis}
In this section, a generalized dynamic $\mathcal{L}_{2}-\mathcal{L}_{\infty}$
filtering method with guaranteed disturbance attenuation level $\mu$ is proposed.

\emph{\textbf{Theorem 1.} Consider the Lipschitz nonlinear system
$\left(\Sigma_{s} \right)$ along with the general filter
$\left(\Sigma_{o} \right)$. The filter error dynamics is
(globally) asymptotically stable with an optimized
$\mathcal{L}_{2}-\mathcal{L}_{\infty}(w \rightarrow e)$ gain, $\mu^{*}$, if there exists
scalars $\zeta>0$, $\epsilon>0$ and $\alpha>0$, and matrices
$C_{F}$, $P_{1}$, $P_{2}$, $G_{1}$, $G_{2}$ and $\mathcal{E}_{3}$, such that
the following optimization problem has a solution.}
\begin{align}
&\hspace{.5cm} \min (\zeta) \notag\\
&\Xi_{1}=\left[
           \begin{array}{ccc}
             \Pi_{1} & \Pi_{2} & \Pi_{3} \\
             \star   & -\epsilon I & 0 \\
             \star   & \star   & -\zeta I \\
           \end{array}
         \right]<0\label{LMI1}\\
&\Xi_{2}=\left[
   \begin{array}{ccccc}
     -\mathbf{E}^{T}P_{1} & C_{F}^{T} & \alpha\gamma I & -C_{F}^{T}H & 0\\
     \star & -\frac{1}{3}I & 0 & 0 & 0\\
     \star & \star & -\frac{1}{3}I & 0 & 0 \\
     \star & \star & \star & \Lambda_{3} & \alpha\gamma I\\
     \star & \star & \star & \star & -\frac{1}{3}I\\
   \end{array}
 \right]<0\label{LMI2}\\
&\Xi_{3}=\left[
   \begin{array}{cc}
     \alpha I & \mathcal{E}_{3}\\
     \star & \alpha I  \\
   \end{array}
 \right]>0\label{LMI3}
\end{align}
\begin{align}
&\Xi_{4}=\left[
           \begin{array}{cc}
             I & I-P_{1}^{T} \\
             \star & I \\
           \end{array}
         \right]>0\label{non-sin}\\
&\mathbf{E}^{T}P_{1}=P_{1}^{T}\mathbf{E}\geq 0 \label{E1}\\
&\mathbf{E}^{T}P_{2}=P_{2}^{T}\mathbf{E}\geq 0 \label{E2}
\end{align}
\emph{where the elements of $\Xi_{1}$ and $\Xi_{2}$ are as defined in the following,
$\Lambda_{1}=G_{1}+G_{1}^{T}+\gamma^{2} I$,
$\Lambda_{2}=A^{T}P_{2}+P_{2}A+\gamma^{2} I+\epsilon N^{T}N$, $\Lambda_{3}=H^{T}H- \mathbf{E}^{T}P_{2}$,
\begin{align}
&\Pi_{1}=\left[
           \begin{array}{cc}
             \Lambda_{1} & G_{2}C \\
             \star & \Lambda_{2}  \\
           \end{array}
         \right], \Pi_{2}=\left[
                            \begin{array}{cc}
                              0 & G_{2}M_{2} \\
                              0 & P_{2}M_{1} \\
                            \end{array}
                          \right], \notag\\
&\Pi_{3}=\left[
           \begin{array}{cccc}
             0 & G_{2} & P_{1}\mathcal{E}_{1} & P_{1}\mathcal{E}_{2} \\
             P_{2} &0 & 0 & 0 \\
           \end{array}
         \right].\notag
\end{align}
Once the problem is solved:}
\begin{align}
A_{F}&=P^{-1}_{1}G_{1}, \ B_{F}=P^{-1}_{1}G_{2}\label{AF-BF}\\
C_{F} \ &\text{is directly obtained},\ \mu^{*} \triangleq \min(\mu) = \sqrt{\zeta^{*}}.\notag
\end{align}
\textbf{Proof:} Consider the following Lyapunov function candidate
\begin{align}
V(\xi(t))=\xi^{T}\mathbf{\widetilde{E}}^{T}P\xi.
\end{align}
To prove the stability of the filter error dynamics, we employ the
well-established generalized Lyapunov stability theory as discussed
in \cite{He_Wang}, \cite{Masubuchi} and \cite{Ishihara} and the
references therein. The generalized Lyapunov stability theory is
mainly based on an extended version of the well-known LaSalle's
invariance principle for descriptor systems. Based on this theory,
the above function along with the conditions \eqref{E1} and
\eqref{E2} is a generalized Lyapunov function (GLF) for the system
$\left(\Sigma_{e} \right)$ where $P=diag(P_{1},P_{2})$. In fact, it
can be shown that $V(\xi(t))=0$ if and only if
$\mathbf{\widetilde{E}}\xi=0$  and positive
elsewhere \cite[Ch. 2]{He_Wang}. Now, we calculate the derivative of $V$ along the
trajectories of $\left(\Sigma_{e} \right)$. We have
\begin{align}
\dot{V}&=\dot{\xi}^{T}\mathbf{\widetilde{E}}^{T} P\xi+\xi^{T}
\mathbf{\widetilde{E}}^{T}P\dot{\xi}=2\xi^{T}(\widetilde{A}+\Delta
\widetilde{A})^{T}P\xi+2\xi^{T}PS_{1}\Omega+2\xi^{T}P\widetilde{B}w.\label{Vdot}
\end{align}
Now, we define
\begin{equation}
J\triangleq V-\int^{\infty}_{0}\left(\mu^{2} w^{T}w\right) dt.
\end{equation}
Therefore,
\begin{equation}
J<\int^{\infty}_{0}\left(\dot{V}-\mu^{2} w^{T}w\right) dt.
\end{equation}
So a sufficient condition for $J\le 0$ is that
\begin{equation}
\forall \ t\in[0,\infty),\hspace{5mm} \dot{V}-\mu^{2} w^{T}w\le 0\label{J1}.
\end{equation}
Thus, using Lemma 1,
\begin{align}
\dot{V}-\mu^{2} w^{T}w \leq& 2\xi^{T}(\widetilde{A}+\Delta \widetilde{A})^{T}P\xi+2\xi^{T}PS_{1}\Omega+2\xi^{T}P\widetilde{B}w-\mu^{2} w^{T}w \notag\\
\leq&\xi^{T}[2(\widetilde{A}+\Delta\widetilde{A})^{T}P+PS_{1}^{T}S_{1}P]\xi+2\xi^{T}P\widetilde{B}w+\Omega^{T}\Omega \notag\\
\leq&\xi^{T}[2(\widetilde{A}+\Delta\widetilde{A})^{T}P+PS_{1}^{T}S_{1}P]\xi+2\xi^{T}P\widetilde{B}w+\Omega^{T}\Omega-\mu^{2} w^{T}w\label{Vdot4}.
\end{align}
Knowing that $\|\Omega(\xi)\| \leq \|\Gamma\xi\|$, we have,
\begin{align}
\Omega^{T}\Omega \leq \gamma^{2} \xi^{T}\xi.\label{Vdot1}
\end{align}
On the other hand,
\begin{align}
\Delta \widetilde{A}&=\left[
                                   \begin{array}{cc}
                                     0 & B_{F}\Delta C \\
                                     0 & \Delta A \\
                                   \end{array}
                                 \right]=\left[
                                           \begin{array}{cc}
                                             0 & B_{F}M_{2}FN \\
                                             0 & M_{1}FN \\
                                           \end{array}
                                         \right]=\left[
                                           \begin{array}{cc}
                                             0 & B_{F}M_{2}\\
                                             0 & M_{1}\\
                                           \end{array}
                                         \right]F\left[
                                                   \begin{array}{cc}
                                                     0 & 0 \\
                                                     0 & N \\
                                                   \end{array}
                                                 \right]\triangleq \widetilde{M_{1}}F\widetilde{N}\label{DAT}.
\end{align}
Therefore, based on \eqref{Vdot1} and using Lemma 2 we can write
\begin{align}
\dot{V}-\mu^{2} w^{T}w <
\xi^{T}[&\widetilde{A}^{T}P+P\widetilde{A}+\epsilon\widetilde{N}^{T}N+\epsilon_{1}^{-1}P\widetilde{M}_{1}\widetilde{M}_{1}P\notag\\
&+PS_{1}S_{1}^{T}P+\gamma^{2}]\xi+2\xi^{T}P\widetilde{B}w-\mu^{2} w^{T}w.\label{Vdot3}
\end{align}
Now, a sufficient condition for \eqref{J1} is that the right hand
side of \eqref{Vdot3} be negative definite. Using Schur complements,
this is equivalent to the following LMI. Note that having $w=0$,
\eqref{Vdot} is already included in \eqref{Vdot4} and consequently
in \eqref{Vdot3}.
{\footnotesize \begin{align} &\left[
  \begin{array}{cccc}
    \widetilde{A}^{T}P+P\widetilde{A}+\gamma^{2} I+\epsilon\widetilde{N}^{T}\widetilde{N}
    & P\widetilde{M}_{1} & PS_{1} & P\widetilde{B} \\
    \star & -\epsilon I & 0 & 0 \\
    \star & \star & -I & 0 \\
    \star & \star & \star & -\zeta I \\
  \end{array}
\right]<0.\notag
\end{align}}
Substituting from \eqref{DAT}, having
$P=diag(P_{1},P_{2})$, defining change of variables $G_{1}\triangleq
P_{1}A_{F}$, $G_{2}\triangleq P_{1}B_{F}$ and $\zeta = \mu^{2}$, and using Schur
complements, the LMI \eqref{LMI1} is obtained.\\
In the next step, we establish the inequality $e^{T}e < V$. We have
$e^{T}e=\xi^{T}\widetilde{C}^{T}\widetilde{C}\xi+2\xi^{T}\widetilde{C}^{T}S_{2}\Omega+\Omega^{T}S_{2}^{T}S_{2}\Omega$.
Therefore,
$e^{T}e \leq 3\xi^{T}\widetilde{C}^{T}\widetilde{C}\xi+3\Omega^{T}S_{2}^{T}S_{2}\Omega$,
while $\Omega^{T}S_{2}^{T}S_{2}\Omega \leq \|S_{2}^{T}S_{2}\| \Omega^{T}\Omega = \|\mathcal{E}_{3}^{T}\mathcal{E}_{3}\| \Omega^{T}\Omega$.
Without loss of generality, we assume that there is a scalar $\alpha$ such that $\|\mathcal{E}_{3}\mathcal{E}_{3}^{T}\|< \alpha^{2}$
(i.e. $\|\mathcal{E}_{3}\|< \alpha$), where $\alpha>0$ is an unknown variable.
Thus, we need to have
$e^{T}e \leq 3\xi^{T}[\widetilde{C}^{T}\widetilde{C}+\alpha^{2}\gamma^{2}I]\xi < \xi^{T}\mathbf{\widetilde{E}}^{T}P\xi$,
which by means of Schur complements is equivalent to the LMI \eqref{LMI2}.
LMI \eqref{LMI3} is equivalent to the condition $\|\mathcal{E}_{3}\|< \alpha_{1}$.
So, based on the above, we have
\begin{align}
\forall \ t \ \ e^{T}(t)e(t) \leq \mu^{2} \int^{\infty}_{0}\left(w^{T}(t)w(t) dt\right),\notag
\end{align}
Therefore, $\|e(t)\|_{\infty} \leq \mu \|w(t)\|_{2}$.
Note that neither $P_{1}$ nor $P_{2}$ are necessarily positive
definite. However, in order to find $A_{F}$ and $B_{F}$ in
\eqref{AF-BF}, $P_{1}$ must be invertible. Since we are
using the spectral matrix norm (matrix 2-norm) throughout this
paper, based on Lemma 3, a sufficient condition for nonsingularity
of $P_{1}$ is that $\|I-P_{1}\|=\sigma_{max}(I-P_{1})<1$. This is
equivalent to $I-(I-P_{1})^{T}(I-P_{1})>0$. Thus, using Schur's
complement, LMI \eqref{non-sin} guarantees the nonsingularity of $P_{1}$.
$\blacksquare$

\emph{\textbf{Remark 1.} The proposed LMIs are linear in $\alpha$
and $\mu$. Thus, either can be a fixed
constant or an optimization variable. Given this, it
may be more realistic to have a combined performance index. This
leads to a multiobjective convex optimization problem optimizing
both $\alpha$ and $\mu$, simultaneously. See \cite{Abbaszadeh3} and
\cite{Abbaszadeh5} for details and examples of multiobjective
optimization approach to filtering for other classes
of nonlinear systems.}

\begin{figure}[!h]
  \centering
  \includegraphics[trim= 3mm 80mm 3mm 15mm, clip, width=.95\textwidth]{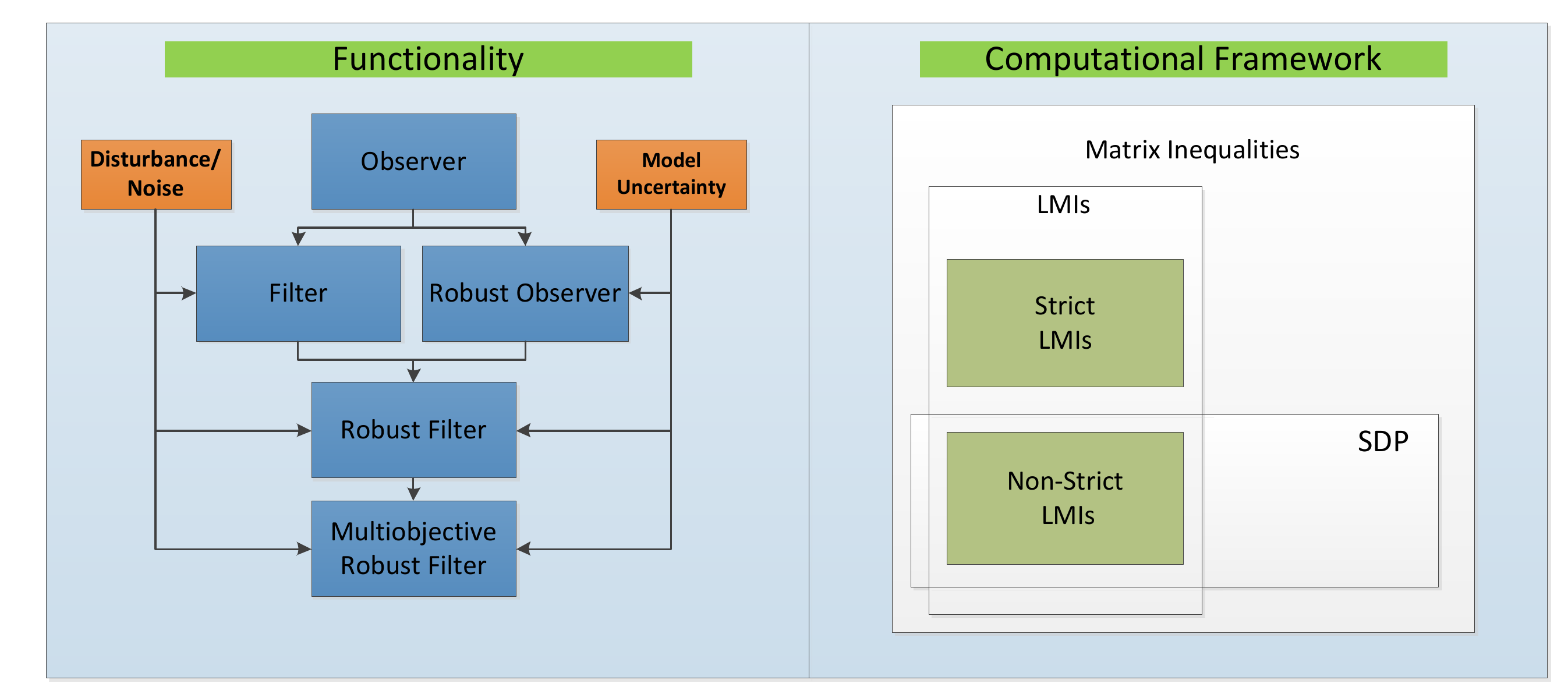}\\
  \caption{State estimation functionality and computational framework}\label{classify_fig}
\end{figure}

Figure \ref{classify_fig} shows a classification of the estate estimators in terms of their functionality,
and the computational frameworks used. Next section will elaborate
further on the differences between strict LMIs and semidefinite programming (SDP).

Note that $\mathcal{E}_{1}$ and $\mathcal{E}_{2}$ are not
optimization variables. They are \emph{apriory} fixed constant
matrices that determine the structure of the filter while
$\mathcal{E}_{3}$ can be either a fixed gain or an optimization variable.

\section{Converting SDP into strict LMIs}
Due to the existence of equalities and non-strict inequalities in
\eqref{E1} and \eqref{E2}, the optimization problem of Theorem 1 is
not a convex \emph{strict} LMI Optimization and instead it is a
Semidefinite Programming (SDP) with quasi-convex solution space. The
SDP problem proposed in Theorem 1 can be solved using freely
available packages such as YALMIP \cite{YALMIP} or SeDuMi \cite{SeDuMi}. However, in order to use the
numerically more efficient Matlab strict LMI solver, in this section we convert the SDP
problem proposed in Theorem 1 into a strict LMI optimization
problem through a smart transformation. We use a similar approach as
used in \cite{Uezato} and \cite{Lu_G3}. Let $\mathbf{E}_{\bot} \in
\mathbb{R}^{(n-s)\times n}$ be the orthogonal complement of
$\mathbf{E}$ such that $\mathbf{E}_{\bot}\mathbf{E}=0$ and
$\text{rank}(\mathbf{E}_{\bot})=n-s$. The following corollary gives
the strict LMI formulation.

\emph{\textbf{Corollary 1.} Consider the Lipschitz nonlinear system
$\left(\Sigma_{s} \right)$ along with the general filter
$\left(\Sigma_{o} \right)$. The filter error dynamics is
(globally) asymptotically stable with an optimized
$\mathcal{L}_{2}-\mathcal{L}_{\infty}(w \rightarrow e)$ gain, $\mu^{*}$, if there exists a
scalars $\zeta>0$, $\epsilon>0$ and $\alpha >0$, and matrices
$C_{F}$, $X_{1}>0$, $X_{2}>0$, $Y_{1}$, $Y_{2}$, $G_{1}$,
$G_{2}$ and $\mathcal{E}_{3}$ such that the following LMI optimization problem has a
solution.}
\begin{align}
&\hspace{.3cm} \min (\zeta) \notag\\
&\Xi_{1}<0, \ \Xi_{2}>0, \ \Xi_{3}>0 \notag\\
&\Xi_{4}=\left[
           \begin{array}{cc}
             I & I-P_{1}^{T} \\
             \star & I \\
           \end{array}
         \right]>0,\notag
\end{align}
\emph{where, $\Xi_{1}$, $\Xi_{2}$, $\Xi_{3}$ and $\Xi_{4}$ are as in
Theorem 1 with}
\begin{align}
P_{1}&=X_{1}\mathbf{E}+\mathbf{E}_{\bot}^{T}Y_{1},
P_{2}=X_{2}\mathbf{E}+\mathbf{E}_{\bot}^{T}Y_{2}\ \label{P1-P2}.
\end{align}
\emph{Once the problem is solved:}
\begin{align}
A_{F}&=P^{-1}_{1}G_{1}=(X_{1}\mathbf{E}+\mathbf{E}_{\bot}^{T}Y_{1})^{-1}G_{1},
B_{F}=P^{-1}_{1}G_{2}=(X_{1}\mathbf{E}+\mathbf{E}_{\bot}^{T}Y_{1})^{-1}G_{2},\ \label{AF2-BF2}\\
C_{F} \ &\text{is directly obtained}, \ \mu^{*} \triangleq \min(\mu) = \sqrt{\zeta^{*}}.\notag
\end{align}
\textbf{Proof:} We have
$\mathbf{E}^{T}P_{1}=\mathbf{E}^{T}(X_{1}\mathbf{E}+\mathbf{E}^{T}_{\perp}Y)=\mathbf{E}^{T}X_{1}\mathbf{E}$.
Since $X_{1}$ is positive definite, $\mathbf{E}^{T}X_{1}\mathbf{E}$
is always at least positive semidefinite (and thus symmetric), i.e.
$\mathbf{E}^{T}P_{1}=P_{1}^{T}\mathbf{E}\geq 0$. Similarly, we have
$\mathbf{E}^{T}P_{2}=P_{2}^{T}\mathbf{E}=\mathbf{E}^{T}X_{2}\mathbf{E}\geq 0$.
Therefore, the two conditions \eqref{E1} and \eqref{E2} are
included in \eqref{P1-P2}.
Now suppose $\widetilde{X}=diag (X_{1},X_{2})$ and
$P=diag(P_{1},P_{2})$. We have
\begin{align}
V=\xi^{T} \mathbf{\widetilde{E}}^{T}P
\xi=\xi^{T}\mathbf{\widetilde{E}}^{T}\widetilde{X}\mathbf{\widetilde{E}}\xi.
\end{align}
Since $X_{1}$ and $X_{2}$ are positive definite, so is
$\widetilde{X}$. Hence, $V$ is always greater than zero and vanishes
if and only if $\mathbf{\widetilde{E}}\xi=0$. Thus, the
transformations \eqref{P1-P2} preserve the legitimacy of
$V$ as a generalized Lyapunov function for the filter error
dynamics. The rest of the proof is the same as the proof of Theorem
1. $\blacksquare$

\emph{\textbf{Remark 2.} The beauty of above result is that with a smart
change of variables the quasi-convex semidefinite programming problem
is converted into a convex strict LMI optimization without any
approximation. Although theoretically, the two problems are
equivalent, numerically the strict LMI optimization problem can be
solved more efficiently. Note that by replacing $P_{1}$ and $P_{2}$
from \eqref{P1-P2} into $\Xi_{1}$ and solving the LMI
optimization problem of Corollary 1, the matrices $X_{1}$, $X_{2}$,
$Y_{1}$ and $Y_{2}$ are directly obtained. Then, having the
nonsingularity of $P_{1}$ guaranteed, the two matrices $A_{F}$ and
$B_{F}$ are obtained as given in \eqref{AF2-BF2},
respectively.}

\section{Numerical Example}
Consider a system of class $\Sigma_{s}$ as
\begin{align}
\left[
      \begin{array}{cc}
      2 & 3 \\
      4 & 6 \\
      \end{array}
\right]\left[
         \begin{array}{c}
           \dot{x}_{1} \\
           \dot{x}_{2} \\
         \end{array}
       \right]&=\left[
    \begin{array}{cc}
      1 & 12 \\
      -6 & -15 \\
    \end{array}
  \right]\left[
         \begin{array}{c}
           x_{1} \\
           x_{2} \\
         \end{array}
       \right]+\frac{1}{2}\left[
        \begin{array}{c}
           \sin x_{2} \\
           \sin x_{1} \\
        \end{array}
      \right]\notag\\
y&= \left[
     \begin{array}{cc}
       1 & 0 \\
     \end{array}
   \right]\left[
         \begin{array}{c}
           x_{1} \\
           x_{2} \\
         \end{array}
       \right]\notag.
\end{align}
We assume the uncertainty and disturbances matrices as follows:
\begin{align}
M_{1}&=\left[
            \begin{array}{cc}
               0.1 & 0.1 \\
               -0.2 & 0.15 \\
            \end{array}
       \right], \ B=\left[
                          \begin{array}{c}
                            1 \\
                            1 \\
                          \end{array}
                     \right], \ N=\left[
                                     \begin{array}{cc}
                                        0.1 & 0 \\
                                        0 & 0.1 \\
                                     \end{array}
                                   \right], \
M_{2}=\left[
         \begin{array}{cc}
           -0.25 & 0.25 \\
         \end{array}
       \right], \ \ D=0.2 \notag.
\end{align}
The system is globally Lipschitz with $\gamma=0.5$.
Now, we design a filter with dynamic structure. Therefore,
we have $\mathcal{E}_{1}=I$ and $\mathcal{E}_{2}=0$.
Using Corollary 1 with $H=0.5I_{2}$, a robust
$\mathcal{L}_{2}-\mathcal{L}_{\infty}$ dynamic filter is obtained as:
\begin{align}
A_{F}&=\left[
        \begin{array}{cc}
        -34.4678 & -19.7142\\
          2.0046 & -28.9571\\
        \end{array}
      \right], \ B_{F}=\left[
         \begin{array}{c}
           1.9586 \\
           0.7948 \\
         \end{array}
       \right], \ C_{F}=\left[
         \begin{array}{cc}
           -0.0111 & -0.0071\\
           -0.0018 & -0.0197\\
         \end{array}
       \right]\notag\\
\epsilon &=1.6437 ,\ \ \ \alpha=4.9876, \ \mu^{*}=0.1453\notag.
\end{align}
As mentioned earlier, in order to simulate the system,
we need consistent initial conditions. Matrix $\mathbf{E}$
is of rank $1$, thus, the system has $1$ differential equation and $1$
algebraic constraint. The system is currently in the \emph{implicit} descriptor form.
In order to extract the algebraic constraint, we can convert the system
into \emph{semi-explicit} differential algebraic. The matrix $\mathbf{E}$ can be decomposed as:
\begin{align}
\mathbf{E}=\left[
      \begin{array}{cc}
      2 & 3 \\
      4 & 6 \\
      \end{array}
\right]=S \left[
      \begin{array}{cc}
      1 & 0 \\
      0 & 0 \\
      \end{array}
\right]T,\ \ S = \left[
      \begin{array}{cc}
      1 & 0 \\
      2 & 1 \\
      \end{array}
\right],\ T = \left[
      \begin{array}{cc}
      3 & 2 \\
      0 & 1 \\
      \end{array}\right].\notag
\end{align}
Now, with the change of variables $\bar{x}=Tx$, the state equations in the original
system are rewritten in the semi-explicit form as follows:
\begin{align}
\left[
      \begin{array}{cc}
      1 & 0 \\
      0 & 0 \\
      \end{array}
\right]\left[
         \begin{array}{c}
           \dot{\bar{x}}_{1} \\
           \dot{\bar{x}}_{2} \\
         \end{array}
       \right]=&\left[
    \begin{array}{cc}
      \frac{1}{3} & \frac{34}{3} \\
      -\frac{8}{3} & -\frac{101}{3} \\
    \end{array}
  \right]\left[
         \begin{array}{c}
           \bar{x}_{1} \\
           \bar{x}_{2} \\
         \end{array}
       \right]+\left[
       \begin{array}{cc}
      \frac{1}{2} & 0 \\
      -1 & \frac{1}{2} \\
    \end{array}
    \right]\left[
        \begin{array}{c}
           \sin \bar{x}_{2} \\
           \sin(\frac{1}{3}\bar{x}_{1}-\frac{2}{3}\bar{x}_{2}) \\
        \end{array}
      \right].\notag
\end{align}
So, the system is clearly decomposed into differential and algebraic parts.
The second equation in the above which is:
\begin{align}
-\frac{8}{3}\bar{x}_{1}-\frac{101}{3}\bar{x}_{2}-sin \bar{x}_{1}+\frac{1}{2}sin(\frac{1}{3}\bar{x}_{1}-\frac{2}{3}\bar{x}_{2})=0,\notag
\end{align}
is the algebraic equation which must be satisfied by the initial conditions. A set of consistent initial conditions satisfying the above equation is found as $\bar{x}_{1}(0)=-38.1034,\ \bar{x}_{2}(0)=3.0014$ which corresponds to $x_{1}(0)=-14.7020,\ x_{2}(0)=3.0014$ which in turn corresponds to $z_{1}(0)=-7.3510,\ z_{2}(0)=1.5007$, where $z=Hx$. Similarly, we find another set of consistent initial conditions for simulating the designed filter. Note that the introduced change of variables is for clarification purposes only to reveal the algebraic constraint which is \emph{implicit} in the original equations which facilitates calculation of consistent initial conditions, and is not required in the filter design algorithm. Consistent initial conditions could also be calculated using the original equations and in fact, most DAE solvers contain a built-in mechanism for consistent initialization using the descriptor form directly. Figure \ref{Fig2} shows the simulation results of $z$ and $z_{F}$ in the absence of disturbance where $z_{F}$ is the output of the filter as in \eqref{observer1}.

\newcommand{\goodgap}{%
\hspace{\subfigtopskip}%
\hspace{\subfigbottomskip}}
\begin{figure}%
\centering
\subfigure[The nominal system]{\label{Fig2}\includegraphics[trim= 42mm 80mm 42mm 80mm,clip=true,width=.45\textwidth]{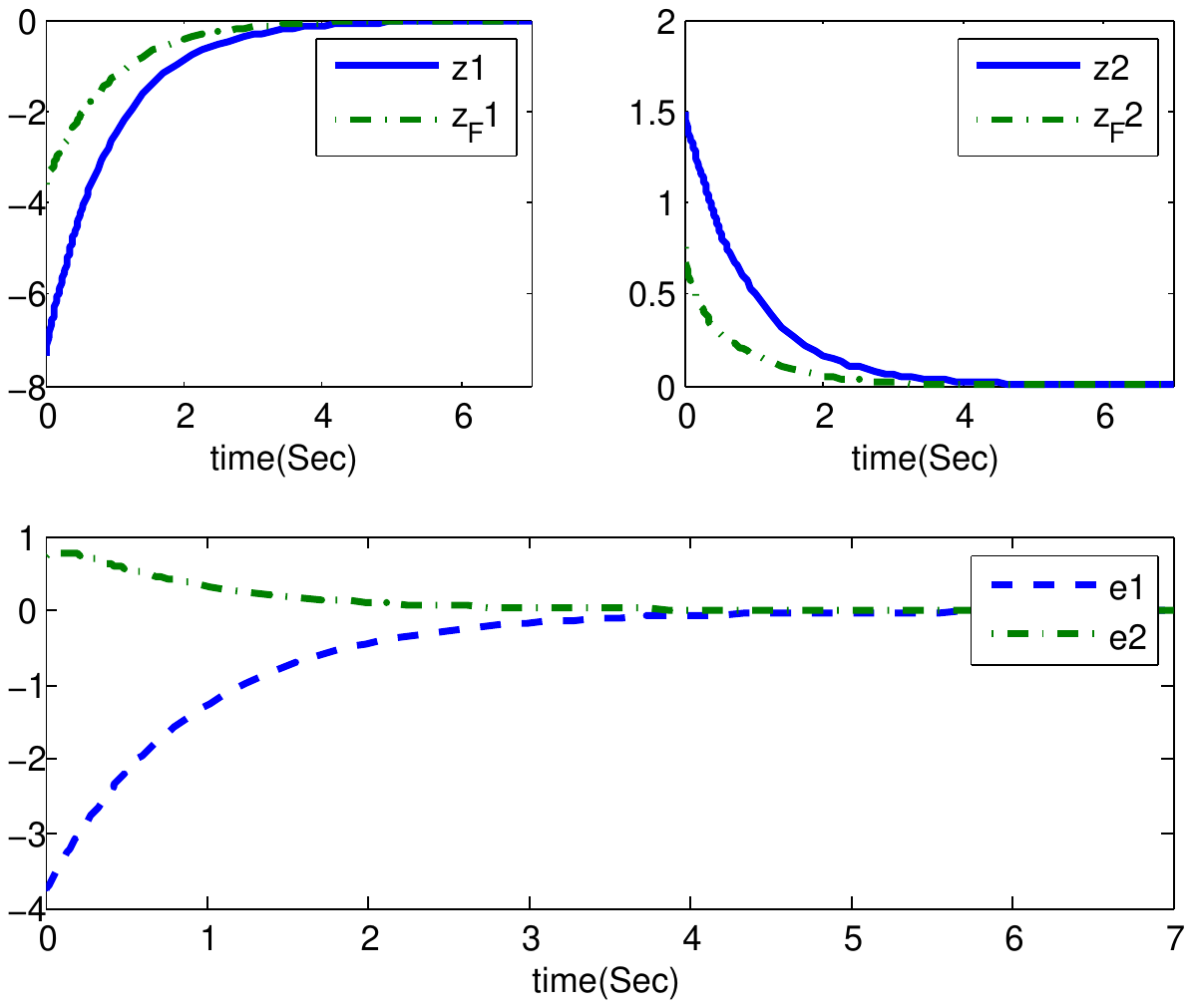}}\goodgap
\subfigure[The disturbed system]{\label{Fig3}\includegraphics[trim= 42mm 80mm 42mm 80mm,clip=true,width=0.45\textwidth]{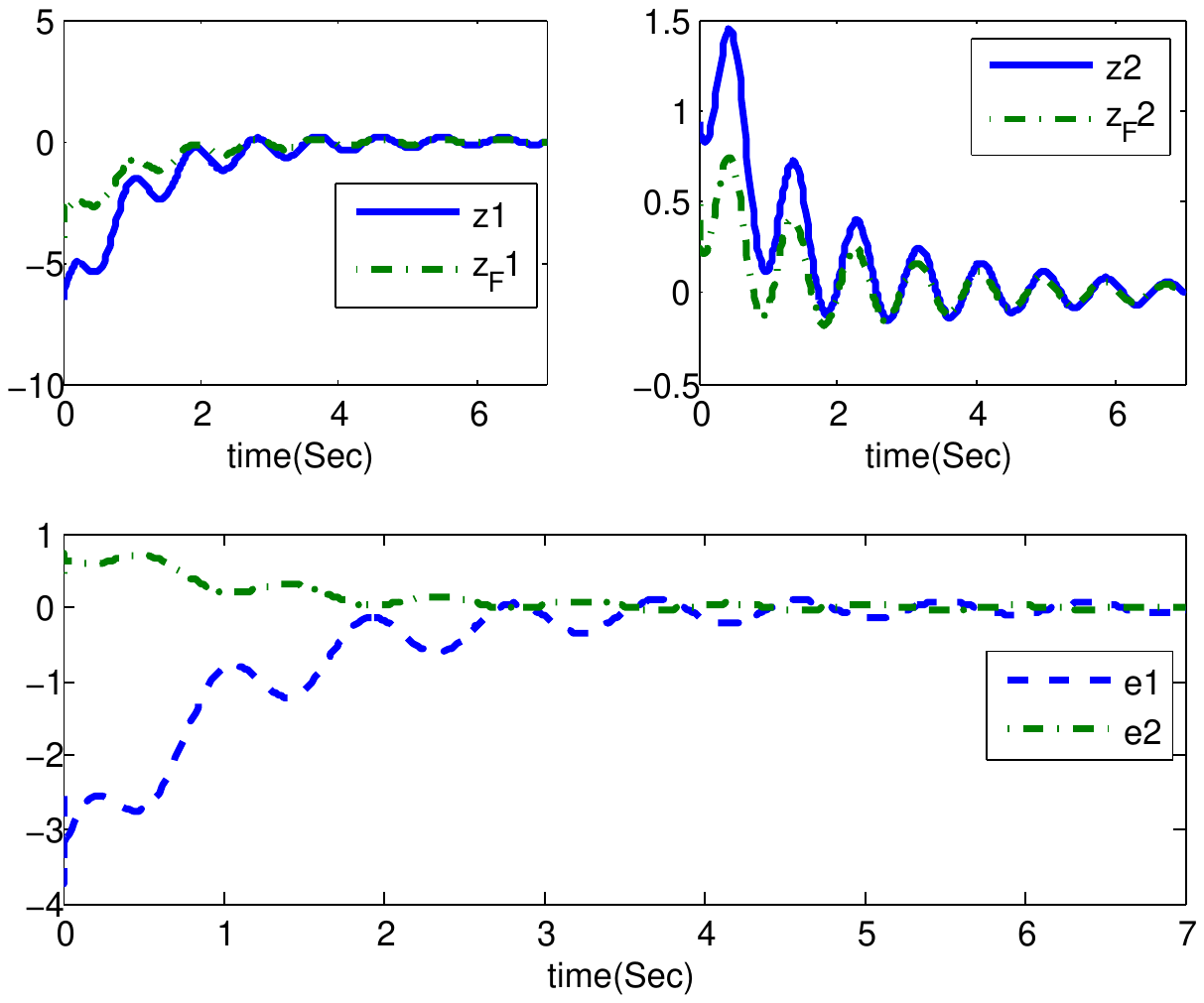}}\\
\caption{Simulations results of the descriptor system and the
           $\mathcal{L}_{2}-\mathcal{L}_{\infty}$ filter.}
\label{3figs}
\end{figure}

Now suppose an unknown $\mathcal{L}_{2}$ exogenous disturbance signal is affecting the system as $w(t)=30\exp(-\frac{t}{3})\cos(7t)$.
Figure \ref{Fig3} shows the simulation results of $z$ and $z_{F}$ in the presence of disturbance.
As expected, in the presence of disturbance, the observer filter error does not converge to zero but it is
kept in the vicinity of zero such that the norm bound $\|e\|_{\infty} \le \mu \|w\|_{2}$ is satisfied.
The designed filter guarantees $\mu$ to be at most $0.1453$. The actual value of $\mu$ for this simulation is $0.0313$.


\section{Conclusion}

A new nonlinear $\mathcal{L}_{2}-\mathcal{L}_{\infty}$ dynamical filter design method for a
class of nonlinear descriptor uncertain systems is proposed through
semidefinite programming and strict LMI optimization.
The proposed dynamical structure has more degree of freedom
than the conventional static-gain filters and is capable of robustly
stabilizing the filter error dynamics for some of those systems for which an
static-gain filter can not be found. The
achieved $\mathcal{L}_{2}-\mathcal{L}_{\infty}$ filter guarantees asymptotic stability of
the error dynamics and is robust against time-varying parametric uncertainty.


\bibliographystyle{plain}
\bibliography{References}


\end{document}